\magnification 1200
\centerline {\bf Quantum Macrostatistical Picture of
Nonequilibrium Steady States}
\vskip 0.5cm
\centerline {{\bf by Geoffrey L. Sewell}\footnote{*}{E-mail
address g.l.sewell@qmul.ac.uk}}
\vskip 0.5cm
\centerline {\bf Department of Physics, Queen Mary, University
of London}
\vskip 0.2cm
\centerline {\bf Mile End Road, London E1 4NS, UK}
\vskip 1cm\noindent
{\bf Abstract.} We apply our quantum macrostatistical treatment
of irreversible processes [18] to prove that, in nonequilibrium
steady states, (a) the hydrodynamical observables execute a
generalised Onsager-Machlup [12] process and (b) the spatial
correlations of these observables are generically of long range.
The key assumptions behind these results are a nonequilibrium
version of Onsager regression hypothesis [11], together with
certain hypotheses of chaoticity and local equilibrium for
hydrodynamical fluctuations.
\vskip 1cm
\centerline {\bf 1. Introduction}
\vskip 0.3cm
It is now well appreciated that a key problem in the statistical
mechanics of irreversible processes is the characterisation of
nonequilibrium steady states [2,6,13], and a number of
different approaches have been made to this problem. Some have
employed rather general strategies, based, for example, on
a hypothesis of Anosov dynamics [6,13]; while others comprise
treatments of concrete microscopic stochastic dynamical models
[20,4,1].
\vskip 0.2cm  
A different approach to nonequilibrium statistical mechanics has
been made by the present author in a number of works [16-18]
that are centred on the hydrodynamical observables of quantum
systems. This approach, like Onsager's [11] treatment of the
subject, is designed to form a bridge beween the microscopic and
macroscopic pictures of matter, rather than a deduction of the
latter from the former. Its basic assumptions concern only very
general, model-independent properties of many-particle systems,
and its scope is thus intended to be complementary to that of
works based on microscopic treatments of many-body problems. The
results to which it has led [18] include a mathematical
characterisation of local thermodynamical equilibrium and a
generalisation of Onsager's reciprocity relations to a
regime where the macroscopic dynamics is nonlinear.
\vskip 0.2cm
In this note, we extend the macrostatistical scheme by the
introducation of a chaoticity hypothesis for the fluctuations of
the non-conserved currents associated with the locally conserved
hydrodynamical observables in nonequilibrium steady states. On
this basis we obtain the results that, in these states, 
\vskip 0.2cm\noindent
(a) the fluctuations of the hydrodynamical observables execute
a generalised Onsager-Machlup (OM) process [12]; and
\vskip 0.2cm\noindent
(b) the spatial correlations of these observables are generically
of long range.  
\vskip 0.2cm\noindent
The latter result constitutes a mathematical generalisation of
results previously proved for certain special classical
stochastic models [20,4,1]. At a heuristic level, similar results
concerning long range correlations have also been obtained from
Landau's fluctuating hydrodynamics [8,5].
\vskip 0.2cm
We remark here that the result (b) marks a qualitative difference
between equilibrium and nonequilibrium steady states, since the
hydrodynamical correlations of the former states are generically
of short range\footnote*{Our distinction between \lq long' and
\lq short' range will be expressed in a sharp mathematical form
in
Section 4}, except at critical points.
\vskip 0.5cm
\centerline {\bf 2. The Model}
\vskip 0.3cm
{\it The Quantum Picture.} We take the model to be an
$N$-particle quantum
system, ${\Sigma}$, that occupies an open, bounded,
connected region, ${\Omega}_{N}$, of a $d$-dimensional Euclidean
space $X$ and is coupled at its surface to an array, ${\cal R}$,
of reservoirs. We assume that the particle number density,
${\nu}$, of ${\Sigma}$ is $N$-independent and that
${\Omega}_{N}$ is the dilation by a factor $L_{N}$ of a fixed,
$N$-independent region ${\Omega}$ of unit volume. Thus
${\Omega}_{N}=L_{N}{\Omega}{\equiv}
{\lbrace}L_{N}x{\vert}x{\in}{\Omega}{\rbrace}$ and
$L_{N}=(N/{\nu})^{1/d}$. In a standard way, we represent the
observables and states of ${\Sigma}$ by the self-adjoint
operators and density matrices, respectively, in a separable
Hilbert space ${\cal H}_{N}$. We assume that, as has been
established under rather general conditions [14, 22], the
composite system $({\Sigma}+{\cal R})$ evolves to a unique
steady state, ${\omega}_{N}$, as $t{\rightarrow}{\infty}$. We
denote the expectation value, for this state, of an observable
$A$ by ${\langle}{\omega}_{N};A{\rangle}$. We shall assume that
all interactions are invariant under space translations and
rotations.
\vskip 0.2cm
We assume that ${\Sigma}$ has a finite set of linearly
independent, extensive, conserved observables ${\hat Q}=({\hat
Q}_{1},. \ .,{\hat Q}_{n})$, which intercommute up to surface
effects and are {\it thermodynamically complete} [18] in the
sense that the states corresponding to pure equilibrium phases
are labelled by the expectation values of their global densities
in the limit $N{\rightarrow}{\infty}$. We denote by $s(q)$ the
equilibrium entropy density corresponding to the value $q$ of the
global density of ${\hat Q}$ in this limit.
\vskip 0.2cm
We assume the observables ${\hat Q}$ have locally conserved,
position dependent densities $\bigl({\hat q}_{1}(x),. \ .,
{\hat q}_{n}(x)\bigr):={\hat q}(x)$, and we denote their evolutes
at time $t$, in the Heisenberg picture of the dynamics of the
composite $({\Sigma}+{\cal R})$, by
${\hat q}^{(N)}(x,t){\equiv}{\hat q}_{t}^{(N)}(x)=
\bigl({\hat q}_{1,t}^{(N)}(x),. \ .,{\hat
q}_{n,t}^{(N)}(x)\bigr)$. In
accordance with the standard requirements of quantum field
theories [21], we assume that the fields ${\hat q}_{j,t}^{(N)}$
are operator-valued distributions. Thus, denoting by ${\cal
D}'({\Omega}_{N})$ the space of continuous linear functionals
on the subset of the L. Schwartz space ${\cal D}(X)$ [15] with
support in ${\Omega}_{N}$, we assume that the fields 
${\hat q}_{t}^{(N)}$ are operator-valued elements of 
${\cal D}'({\Omega}_{N})^{n}$. 
\vskip 0.3cm
{\it The Hydrodynamical Picture of ${\Sigma}$.} We assume
that this is given by a continuum mechanical law governing the
evolution of a set of locally conserved classical fields
$q_{t}(x)=\bigl(q_{1,t}(x),. \ .,q_{n,t}(x)\bigr)$, on a
macroscopic space-time scale that we shall presently specify.
Here the fields $q_{j,t}(x)$ represent the densities at position
$x$ and time $t$ of the extensive thermodynamical variables of
${\Sigma}$. In general, their dynamics is given by a
classical equation of motion of the form
$${{\partial}q_{t}(x)\over {\partial}t}={\cal F}(q_{t};x),$$
the right hand side representing the value of a functional ${\cal
F}$ of $q_{t}$ at the position $x$. For simplicity we base our
treatment here on the case where this equation corresponds to a
nonlinear diffusion and thus takes the form
$${{\partial}q_{t}\over {\partial}t}=
{\nabla}.\bigl(K(q_{t}){\nabla}q_{t}\bigr),\eqno(1)$$
subject to certain stationary, spatially varying boundary
conditions determined by the reservoirs ${\cal R}$. Here and
elsewhere $q_{t}$ is a function of position,
and $K(q_{t})$ is an $n$-by-$n$ matrix $[K_{jk}(q_{t})]$, whose
action on ${\nabla}q_{t}$ is given by standard matrix
multiplication. We shall presently specify the relationship
between $q_{t}$ and the quantum field ${\hat q}_{t}^{(N)}$. 
\vskip 0.2cm
We assume that the unit of the macroscopic length scale is the
distance $L_{N}$, introduced above; and correspondingly, since
Eq. (1) is invariant under the scale transformations
$x{\rightarrow}{\lambda}x, \
t{\rightarrow}{\lambda}^{2}t$, we take the macroscopic time scale
to be $L_{N}^{2}$. Thus, Eq. (1) represents a phenomenolgical
dynamics in the bounded spatial region ${\Omega}$ on this time
scale. We assume that that equation admits a unique stationary
solution, $q_{t}=q$, subject to the imposed boundary
conditions.  We assume throughout that the system is in a single
phase region wherein both $K(q)$ and the entropy density $s(q)$
are smooth functions of $q$ as this variable runs over the range
of $q(x)$ for $x{\in}{\Omega}$.
\vskip 0.2cm
A simple important consequence of Eq. (1) is that, although the
fields $q_{t}$ are locally conserved, the associated currents are
not. This is crucial for our key assumption, formulated in
Section 3, to the effect that the irreversibility of the flow of
$q_{t}$ stems from chaoticity properties of the fluctuations of
these
currents\footnote*{In the case of Navier-Stokes hydrodynamics,
it would be the stress tensor, not the mass current, that would
be the locally non-conserved field, and our chaoticity
assumption would pertain to the fluctuations in this tensor.}.
\vskip 0.3cm 
{\it Relationship between the Quantum and Hydrodynamical
Pictures.} We assume that the classical field $q$ is the
expectation value of ${\hat q}$ for the steady state
${\omega}_{N}$, as represented on the macroscopic length scale
of unit $L_{N}$. Thus,
$$q(x)={\rm lim}_{N\to\infty}{\langle}{\omega}_{N};
{\hat q}(L_{N}x){\rangle} \ {\forall} \
x{\in}{\Omega}.\eqno(2)$$
Correspondingly, since the unit of the macroscopic time scale is
$L_{N}^{2}$, we represent the time-dependent fluctuations of
${\hat q}_{t}(x)
\ \bigl({\equiv}{\hat q}(x,t)\bigr)$ about its steady state
value, on the macroscopic scale, by the field
$${\xi}_{t}^{(N)}(x)=
N^{1/2}\bigl[{\hat q}^{(N)}(L_{N}x,L_{N}^{2}t)
-{\langle}{\omega}_{N};{\hat q}(L_{N}x){\rangle}],\eqno(3)$$
the normalisation parameter $N^{1/2}$ being the canonical one for
fluctuations [7]. 
\vskip 0.5cm
\centerline {\bf 3. The Fluctuations: a Generalised
Onsager-Machlup Process.} 
\vskip 0.3cm
We now assume, on grounds  [16-18] representing the
intercommutativity of the quantum
fields ${\hat q}_{j,t}(x)$ and ${\hat q}_{k,u}(y)$ at macroscopic
space-time separation and the fact that the ratio of macroscopic
to microscopic relaxation times becomes infinite in the limit
$N{\rightarrow}{\infty}$, that the quantum process
${\xi}_{t}^{(N)}$ reduces to a classical stationary Markov
process ${\xi}_{t}$ in this limit, i.e. that
$${\rm lim}_{N\to\infty}{\langle}{\omega}_{N};
{\xi}_{t_{1}}^{(N)}(x_{1}){\otimes}{\xi}_{t_{2}}^{(N)}(x_{2})
.. \ .{\otimes}{\xi}_{t_{m}}^{(N)}(x_{m}){\rangle}=
E\bigl({\xi}_{t_{1}}(x_{1}){\otimes}{\xi}_{t_{2}}(x_{2})
.. \ .{\otimes}{\xi}_{t_{m}}(x_{m})\bigr)$$
$$ \ {\forall} \ x_{1},. \ .,x_{m}{\in}{\Omega}, 
\ t_{1},. \ .,t_{m}{\in}{\bf R}_{+},\eqno(4)$$
where ${\otimes}$ denotes the ${\bf R}^{n}$ tensor product.
Since ${\cal D}'$ spaces are complete, it follows from this
formula that ${\xi}_{t}$ is a distribution-valued random field.
The forward time derivative of ${\xi}_{t}$, as defined by Nelson
[10], is
$$D{\xi}_{t}={\lim}_{h{\rightarrow}+0}h^{-1}
E\bigl({\xi}_{t+h}-{\xi}_{t}{\vert}{\xi}_{t}\bigr),\eqno(5)$$
where $E(.{\vert}{\xi}_{t})$ denotes the conditional expectation
functional, given the field ${\xi}_{t}$.
\vskip 0.3cm
{\it Regression Hypothesis.} We now invoke a version of Onsager's
regression hypothesis [11], to the effect that the regressions
of the fluctuations ${\xi}_{t}$ are governed by the same
dynamical law as the \lq small' perturbations ${\delta}q_{t}$ of
the
macroscopic field $q$. To this end, we start by inferring from
Eq. (1) that the linearized equation of motion for
${\delta}q_{t}$ is
$${{\partial}\over {\partial}t}{\delta}q_{t}(x)=
{\cal L}{\delta}q_{t}(x):=
{\nabla}.\bigl(K(q(x)){\nabla}({\delta}q_{t}(x))
+[K'(q(x)){\delta}q_{t}(x)]{\nabla}q(x)\bigr),\eqno(6)$$
where $K'(q)$ is the derivative of $K(q)$, i.e. its gradient
w.r.t. the variable $q$: thus $[K'(q){\delta}q_{t}]_{jk}=
{\sum}_{l=1}^{n}[{\partial}K_{jk}(q)/{\partial}q_{l}]
{\delta}q_{l,t}$. We assume that the perturbation of $q$ does not
change its boundary conditions, i.e. that ${\delta}q_{t}$
vanishes on the boundary of ${\Omega}$. 
\vskip 0.2cm
We assume that ${\delta}q_{t}$, like the quantum field ${\hat
q}$, is a ${\cal D}'({\Omega})^{n}$-class distribution and that
the operator ${\cal L}$, defined in Eq. (6), is the generator of
a one-parameter semigroup of transformations
${\lbrace}T_{t}{\vert}t{\in}{\bf R}_{+}{\rbrace}$ of 
${\cal D}'({\Omega})^{n}$, as determined by the equation
$${dT_{t}\over dt}={\cal L}T_{t}; \ T_{0}=I;\eqno(7)$$
and thus that the solution of Eq. (6) is
$${\delta}q_{t}=T_{t-s}{\delta}q_{s} \ {\forall} \
t{\geq}s{\geq}0.\eqno(8)$$ 
We assume that the system satisfies the dissipativity condition
that the perturbation ${\delta}q_{t}$ vanishes in the limit
$t{\rightarrow}{\infty}$, i.e. that
$${\lim}_{t{\rightarrow}{\infty}}T_{t}{\phi}=0,\eqno(9)$$
for all elements of ${\phi}$ that vanish on the boundary of
${\Omega}$. 
\vskip 0.2cm
We now assume, as a generalisation of Onsager's
regression hypothesis for equilibrium fluctuations that, for
$t{\geq}s{\geq}0$, the
conditional expectation of ${\xi}_{t}$, given ${\xi}_{s}$, is
$$E\bigl({\xi}_{t}{\vert}{\xi}_{s}\bigr)=
T_{t-s}{\xi}_{s} \ {\forall} \
t{\geq}s{\geq}0.\eqno(10)$$
Hence, by Eqs. (5), (7) and (10),
$$D{\xi}_{t}={\cal L}{\xi}_{t}.\eqno(11)$$
Further, by Eq. (10) and the stationarity of the ${\xi}_{t}$
process, the two-point function
$E\bigl({\xi}_{t}(x){\otimes}{\xi}_{t'}(x')\bigr)$
takes the value 
$(T_{t-t'}{\otimes}I)E\bigl({\xi}(x){\otimes}{\xi}(x')\bigr)$ or
$(I{\otimes}T_{t'-t})E\bigl({\xi}(x){\otimes}
{\xi}(x')\bigr)$ according to whether or not $t{\geq}t'$. Thus,
$$E\bigl({\xi}_{t}(x){\otimes}{\xi}_{t'}(x')\bigr)=$$
$$(T_{t-t'}{\otimes}I)
E\bigl({\xi}(x){\otimes}{\xi}(x')\bigr)
{\theta}(t-t')+(I{\otimes}T_{t'-t})
E\bigl({\xi}(x){\otimes}{\xi}(x')\bigr)
\bigl(1-{\theta}(t-t')\bigr),\eqno(12)$$
where ${\theta}$ is the Heaviside function that takes the value
$1$ or $0$ according to whether or not its argument is non-
negative.
\vskip 0.3cm
{\it Extended Stochastic Process: the Currents.} Recalling that
the fields ${\hat q}_{t}$ are locally conserved observables, we
assume that the above formulation of the stochastic process
${\xi}_{t}$ has a canonical extension to a larger process 
$({\xi}_{t},{\eta}_{t})$, where ${\eta}_{t}$ represents the
fluctuations of the currents\footnote*{Note that it follows from
the local conservation laws that the divergences of these
currents are also classical. Hence, it is irrelevant for our
purposes that the currents themselves might not be classical,
since they enter into our calculations only through
${\nabla}.{\eta}$. In fact, we have obtained the same results by
a fully quantum treatment [19] of the currents, which does not
contain formal conditional expectations such as that in Eq.
(15).}  associated with ${\hat q}_{t}$.
Thus, ${\xi}_{t}$ and ${\eta}_{t}$ conform to the local
conservation law
$${{\partial}{\xi}_{t}\over {\partial}t}+{\nabla}.{\eta}_{t}=0,
\eqno(13)$$
and consequently, by Eqs. (5) and (11),
$${\cal L}{\xi}_{t}=
-E\bigl({\nabla}.{\eta}_{t+0}{\vert}{\xi}_{t}\bigr).\eqno(14)$$ 
We now define
$${\tilde {\eta}}_{t}={\eta}_{t}-
E\bigl({\eta}_{t+0}{\vert}{\xi}_{t}\bigr)\eqno(15)$$
and
$$b_{t}=-{\nabla}.{\tilde {\eta}}_{t}{\equiv}
-{{\partial}\over {\partial}x_{\mu}}{\tilde {\eta}}_{{\mu},t},
\eqno(16)$$
where $x_{\mu}$ and ${\eta}_{{\mu},t}$ are the ${\mu}$'th
Cartesian components of $x$ and ${\tilde {\eta}}_{t}$,
respectively. We note that Eq. (15) may naturally be interpreted
as signifying that $E\bigl({\eta}_{t+0}{\vert}{\xi}_{t}\bigr)$
and  ${\tilde {\eta}}_{t}$ are the secular and residual
stochastic parts, respectively, of ${\eta}_{t}$; and thus Eq.
(16) signifies that $b_{t}$ is minus the divergence of the
stochastic part of the current.
\vskip 0.2cm 
It follows now from Eqs. (13)-(16) that 
$${d{\xi}_{t}\over dt}={\cal L}{\xi}_{t}+b_{t},\eqno(17)$$
which has the {\it form} of a Langevin
equation\footnote*{Strictly speaking, ${\tilde {\eta}}_{t}$ and
$b_{t}$ should be treated as distributions w.r.t. $t$, since it
will presently emerge that each of them corresponds
to a white noise (cf. Eqs. (18) and (29)). Mathematical propriety
can easily be achieved, however, without changing the structure
of the argument, by working with the integral
$w_{t}=\int_{0}^{t}dub_{u}$, which would correspond to a Wiener
process, instead of $b_{t}$.}. Our interpretation of it, however,
will depend on the chaoticity and local equilibrium hypotheses
that we shall presently introduce. At all events, $b_{t}$ is
statistically independent of ${\xi} \ (:={\xi}_{0})$ for $t>0$,
since, by Eqs. (7), (10) and (17),
$$E(b_{t}{\vert}{\xi})={d\over dt}E({\xi}_{t}{\vert}{\xi})-
{\cal L}E({\xi}_{t}{\vert}{\xi})=({d\over dt}-{\cal L})T_{t}{\xi}
=0.$$
Further, by Eq. (17),
$$E\bigl(b_{t}(x){\otimes}b_{t'}(x')\bigr)=
\bigl({{\partial}\over {\partial}t}-{\cal L}{\otimes}I\bigr)
\bigl({{\partial}\over {\partial}t'}-I{\otimes}{\cal L}'\bigr)
E\bigl({\xi}_{t}(x){\otimes}{\xi}_{t'}(x')\bigr),$$
where ${\cal L}'$ is the image of ${\cal L}$, as defined by Eq.
(6), under the transformation $x{\rightarrow}x'$.
Hence, by Eqs. (7) and (12), together with the identity
$d{\theta}(t-t')/dt{\equiv}{\delta}(t-t')$,
$$E\bigl(b_{t}(x){\otimes}b_{t'}(x')\bigr)=  
-\bigr[E\bigl({\cal L}{\xi}(x){\otimes}{\xi}(x')\bigr)
+E\bigl({\xi}(x){\otimes}{\cal L}'{\xi}(x')\bigr)\bigr]
{\delta}(t-t').\eqno(18)$$
The higher order correlation functions of $b_{t}$ will be
governed by the following chaoticity hypothesis.
\vskip 0.3cm
{\it Chaoticity Hypothesis.} This is the hypothesis that the
space-time correlations of the stochastic part of the
non-conserved current fluctuations associated with ${\hat
q}_{t}$, as viewed on the microscopic scale, are of short range.
Since the ratios of the macroscopic to microscopic scales of both
length and time are infinite, this signifies that the space-time
correlations of the currents ${\tilde {\eta}}_{t}(x)$ have zero
range. Further, since the fluctuations of quantum fields with
short range correlations are generally Gaussian in the large
scale limit [7], we assume that ${\tilde {\eta}}_{t}$ is a
Gaussian process. Thus, our chaoticity assumption is that
${\tilde {\eta}}_{t}$ is a Gaussian field with zero range
space-time correlations. Hence, by a standard theorem
on distributions [15; Theorem 35], we can state the chaoticity
hypothesis as follows. 
\vskip 0.3cm
{\it (C) ${\tilde {\eta}}_{t}$ is a stationary Gaussian process
whose two- point function
$E\bigl({\tilde {\eta}}_{{\mu},t}(x){\otimes}
{\tilde {\eta}}_{{\nu},t'}(x')\bigr)$
is a finite linear combination of ${\delta}(x-x'){\delta}(t-t')$
and its derivatives, with coefficients given by generalised
functions of $x$.} 
\vskip 0.3cm
We note here that the Gaussian property of $b_{t}$ implies that
of ${\xi}_{t}$, for the following reasons. Since ${\cal L}$ is
the generator of the semigroup $T$, it follows from Eq.(17) that
$${\xi}_{t+t_{0}}=T_{t+t_{0}}{\xi}+
\int_{0}^{t+t_{0}}dsT_{t+t_{0}-s}b_{s} \ {\forall} \
t,t_{0}{\geq}0.$$
By {\it (C)}, the integral on the r.h.s. of this equation is
Gaussian, while, by Eq. (9), the first term on its r.h.s.
vanishes in the limit $t{\rightarrow}{\infty}$. Hence, as the
stationarity of the process ${\xi}_{t}$ implies that it is
isomorphic with ${\xi}_{t+t_{0}}$, it must be Gaussian. 
\vskip 0.3cm
{\it Equilibrium Conditions.} As a preliminary to the formulation
of local equilibrium conditions for the nonequilibrium steady
state, we first formulate the true equilibrium two-point
functions of the fields ${\xi}_{t}$ and ${\eta}_{t}$. For this
purpose we assume that the equilibrium state is achieved by
arranging the reservoirs ${\cal R}$ so that $q(x)$ is uniform on
the boundary, ${\partial}{\Omega}$, of ${\Omega}$. With this
boundary condition, Eq. (1) has a stationary solution in which
$q(x)$ is uniform throughout ${\Omega}$. Consequently, by our
uniqueness assumption of Section 2, $q(x)$ reduces to a constant,
$q$ in the equilibrium state. Further, recalling our assumption
of Section 2 that the interactions of the system are
translationally and rotationally invariant, we assume
that the corresponding symmetries are not broken in the pure
equilibrium phase and thus that the process $({\xi}_{t}, \
{\tilde {\eta}}_{t})$ is invariant under the space translations
and rotations that are implementable within the confines of
${\Omega}$. We remark here that the limitation in the Euclidean
symmetry imposed by the boundedness of ${\Omega}$ is not serious
from the physical standpoint, since any point of this {\it open}
region corresponds, in the microscopic picture, to one that is
infinitely far from the boundary of the quantum system
${\Sigma}$.
\vskip 0.2cm
Under the above assumptions, a quantum statistical treatment of
the field ${\xi}$ yields the following result for the static two-
point function for ${\xi}$ (cf. [18, Ch. 7, Appendix C]): in
fact,
it corresponds to a thermodynamic limiting
version of the Einstein formula $P={\rm const.}{\rm exp}(S)$.
$$E_{equil}\bigl({\xi}(x){\otimes}{\xi}(x')\bigr)=J(q){\delta}
(x-x'),\eqno(19)$$
where $E_{equil}$ is the equilibrium expectation functional and
$J(q)$ is minus the inverse matrix of the Hessian of the entropy
density function $s(q)$, i.e.
$$J(q)=-[{\partial}^{2}s(q)/{\partial}q_{j}{\partial}q_{k}]^{-1}.
\eqno(20)$$
On the other hand, it follows from the assumptions of
translational and rotational invariance that the two-point
function of ${\tilde {\eta}}_{t}(x)$ takes the form
$$E_{equil}\bigl({\tilde {\eta}}_{{\mu},t}(x)
{\otimes}{\tilde {\eta}}_{{\nu},t'}(x')\bigr)=
G(x-x',t-t'){\delta}_{{\mu}{\nu}},\eqno(21)$$
where $G$ is a scalar w.r.t. the vector space $X$. Hence, the
chaoticity condition $(C)$ reduces here to the following form.
\vskip 0.3cm
{\it $(C)_{equil}$. $G(x-x',t-t')$ is a finite linear
combination, with constant coefficients, of the distribution
${\delta}(x-x'){\delta}(t-t')$ and its derivatives.}
\vskip 0.3cm
In order to pin down the explicit form of $G$ we now note that,
by Eq. (6), the constancy of $q$ implies that ${\cal L}$ is
simply $K(q){\Delta}$, and therefore, by Eqs. (16), (18) and
(19), that
$$E_{equil}\bigl(b_{t}(x){\otimes}b_{t'}(x')\bigr)=
{{\partial}^{2}\over {\partial}x_{\mu}{\partial}x_{\nu}'}
E_{equil}\bigl({\tilde {\eta}}_{{\mu},t}(x){\otimes}
{\tilde {\eta}}_{{\nu},t'}(x')\bigr)$$
$$=2[K(q)J(q)]_{sym}{\Delta}{\delta}(x-x')
{\delta}(t-t'),\eqno(22)$$
where, for any matrix $A, \  [A]_{sym}$ denotes the arithmetic
mean of $A$ and its transpose\footnote*{In fact, it has been
proved in [18] that $K(q)J(q)$ is a symmetric matrix, i.e. that
the Onsager reciprocity relations prevail, subject to the
assumption of microscopic reversibility. Moreover, this result
was extended there to the nonequilibrium situation under the
assumption of a certain local equilibrium hypothesis not employed
in this note. However, we shall not assume this symmetry here,
since it is not needed for our present purposes.}. Hence, by Eqs.
(21) and (22),
$${\Delta}G(x-x',t-t')= 
2[K(q)J(q)]_{sym}{\Delta}{\delta}(x-x'){\delta}(t-t').$$
This equation, together with condition $(C)_{equil}$, fixes the
form of $G$ according to the formula
$$G(x-x',t-t')=2[K(q)J(q)]_{sym}{\delta}(x-x')
{\delta}(t-t'),$$
and consequently, by Eq. (21),
$$E_{equil}\bigl({\tilde {\eta}}_{{\mu},t}(x)
{\otimes}{\tilde {\eta}}_{{\nu},t'}(x')\bigr)=
2[K(q)J(q)]_{sym}{\delta}(x-x'){\delta}(t-t')
{\delta}_{{\mu}{\nu}}.\eqno(23)$$
\vskip 0.2cm
We now note that, for any positive
${\epsilon}$ and any $x_{0}{\in}{\Omega}$ and $t_{0}{\in}
{\bf R}_{+}$, Eqs. (19) and (23) are invariant under the
transformations
$x{\rightarrow}x_{0}+{\epsilon}x, \
x'{\rightarrow}x_{0}+{\epsilon}x', \
{\xi}{\rightarrow}{\epsilon}^{d/2}{\xi}$ and
$x{\rightarrow}x_{0}+{\epsilon}x, \
x'{\rightarrow}x_{0}+{\epsilon}x', \
t{\rightarrow}t_{0}+{\epsilon}^{2}t, \
t'{\rightarrow}t_{0}+{\epsilon}^{2}t', \
{\tilde {\eta}}_{t}{\rightarrow}{\epsilon}^{1+d/2}
{\tilde {\eta}}_{t}$, respectively. Thus they are
equivalent to the following formulae.
$${\epsilon}^{d}E_{equil}\bigl({\xi}(x_{0}+{\epsilon}x){\otimes}
{\xi}(x_{0}+{\epsilon}x')\bigr)=J(q){\delta}
(x-x'),\eqno(24)$$
and
$${\epsilon}^{d+2}E_{equil}
\bigl({\tilde {\eta}}_{{\mu},t_{0}+{\epsilon}^{2}t}
(x_{0}+{\epsilon}x){\otimes}
{\tilde {\eta}}_{{\nu},t_{0}+{\epsilon}^{2}t'}
(x_{0}+{\epsilon}x')\bigr)=2[K(q)J(q)]_{sym}
{\delta}(x-x'){\delta}(t-t').\eqno(25)$$
Evidently, if ${\epsilon}$ is chosen to be \lq small', then these
equations correspond to local conditions concentrated at the
space-time point $(x_{0},t_{0})$.
\vskip 0.3cm
{\it Local Equilibrium Conditions.} This last observation leads
us to propose the following local equilibrium conditions for the
general stationary nonequililibrium situation.
$${\rm lim}_{{\epsilon}{\rightarrow}0}
{\epsilon}^{d}E\bigl({\xi}(x_{0}+{\epsilon}x){\otimes}
{\xi}(x_{0}+{\epsilon}x')\bigr)=J\bigl(q(x_{0})){\delta}
(x-x')\eqno(26)$$
and
$${\rm lim}_{{\epsilon}{\rightarrow}0}{\epsilon}^{d+2}
E\bigl({\tilde {\eta}}_{{\mu},t_{0}+{\epsilon}^{2}t}
(x_{0}+{\epsilon}x){\otimes}
{\tilde {\eta}}_{{\nu},t_{0}+{\epsilon}^{2}t'}
(x_{0}+{\epsilon}x')\bigr)=$$
$$2\bigl[K\bigl(q(x_{0})\bigr)
J\bigl(q(x_{0})\bigr)\bigr]_{sym}
{\delta}(x-x'){\delta}(t-t'){\delta}_{{\mu}{\nu}}.\eqno(27)$$
\vskip 0.3cm
{\it The Generalised Onsager-Machlup Process.} In view of the
chaoticity condition {\it (C)}, it follows immediately from this
last equation that the two-point function for ${\tilde
{\eta}}_{t}$ must take the following form, since the presence of
derivatives of ${\delta}(x-x'){\delta}(t-t')$ would render the
l.h.s. of Eq. (27) divergent.
$$E\bigl({\tilde {\eta}}_{{\mu},t}(x){\otimes}
{\tilde {\eta}}_{{\nu},t'}(x')\bigr)=
2\bigl[K\bigl(q(x)\bigr)J\bigl(q(x)\bigr)\bigr]_{sym}
{\delta}(x-x'){\delta}(t-t'){\delta}_{{\mu}{\nu}}.\eqno(28)$$
It follows from this formula and Eq. (16), the two-point function
for $b_{t}$ is
$$E\bigl(b_{t}(x){\otimes}b_{t'}(x')\bigr)=
-2{\nabla}.\bigl(\bigl[K\bigl(q(x)\bigr)
J\bigl(q(x)\bigr)\bigr]_{sym}
{\nabla}{\delta}(x-x')\bigr){\delta}(t-t').\eqno(29)$$
Hence, by the Gaussian assumption in {\it (C)}, $b_{t}$ is a
white noise and consequently, by Eq. (17), ${\xi}_{t}$ executes
a generalised Onsager-Machlup process [12]. We remark here that
the two-point function for $b_{t}$ given by Eq. (29) is of the
same form as that of the noise term in Landau's fluctuating
hydrodynamics [9].
\vskip 0.5cm
\centerline {\bf 4. Long Range Spatial Correlations.} 
\vskip 0.3cm
The static two-point function for the fluctuation field ${\xi}$
is
$$W(x,x')=E\bigl({\xi}(x){\otimes}{\xi}(x')\bigr)
{\equiv}E\bigl({\xi}_{t}(x){\otimes}{\xi}_{t}(x')\bigr) \ 
{\rm by} \ {\rm stationarity}.\eqno(30)$$
Since the ratio of the macroscopic to microscopic length scale
is infinite, short range correlations on the latter scale reduce
to zero on the former one. Accordingly, we term the range of the
correlations \lq short' or \lq long' according to whether or not
it
reduces to zero in the macroscopic picture. Hence our condition
for long range correlations is simply that the support of
the distribution $W$ does {\it not} lie in the domain
${\lbrace}(x,x'){\in}{\Omega}^{2}{\vert}x=x'{\rbrace}$. The
following Proposition establishes that the spatial correlations
of ${\xi}$ are generically of long range.
\vskip 0.3cm
{\bf Proposition.} {\it Let ${\Phi}$ be the $n$-by-$n$ matrix-
valued functional on the classical field $q$ defined by the
formula
$${\Phi}(q;x)={\Delta}\bigl[K\bigl(q(x)\bigr)
J\bigl(q(x)\bigr)\bigr]_{sym}+
{\nabla}.{\Psi}(q;x)_{sym}\eqno(31),$$
where
$${\Psi}_{jk}(q;x)=\Bigl[{{\partial}\over {\partial}q_{m}(x)}
K_{jl}\bigl(q(x)\bigr)\Bigr]
\bigl[J_{mk}\bigl(q(x)\bigr){\nabla}q_{l}(x)-
J_{lk}\bigl(q(x)\bigr){\nabla}q_{m}(x)\bigr].
\eqno(32)$$ 
Then under the above assumptions, a sufficient
condition for the spatial correlations of ${\xi}$ to be of long
range is that ${\Phi}(q)$ does not vanish.}
\vskip 0.3cm
{\bf Comments.} (1) The Proposition establishes that
the correlations are generically of long range, since the
condition that ${\Phi}(q)$ vanishes can be satified only for
special relationships between the functions $K\bigl(q(x)\bigr)$
and $s\bigl(q(x)\bigr)$. By contrast, the corresponding
correlations for equilibrium states are generically of short
range, except at critical points. A treatment of critical
equilibrium correlations of fluctuation observables is provided
in Ref. [3].
\vskip 0.2cm 
(2) In the particular case of the symmetric
exclusion process [20,4], $n=1, \ d=1, \ K(q)=1, \
s(q)=-q{\ln}q-
(1-q){\ln}(1-q)$ and $q(x)=a+b.x$, where $a$ and $b \ ({\neq}0)$
are constants. Thus, in this case, it follows from Eqs. (20),
(31) and (32) that ${\Phi}(q;x)=-2b^{2}{\neq}0$. Hence, as proved
by other methods in Refs. [20, 4], long range correlations
prevail
in this model.
\vskip 0.3cm
{\bf Proof of Proposition.} Suppose that the spatial correlations
of ${\xi}$ are not of long range, i.e. that the support of $W$
does lie in the domain
${\lbrace}(x,x'){\in}{\Omega}^{2}{\vert}x=x'{\rbrace}$. Then it
follows from a classic theorem on distributions [15; Theorem 35]
that $W(x,x')$ is a finite linear combination of ${\delta}(x-x')$
and its derivatives, with coefficients given by generalised
functions of $x$. Under this assumption, it follows from Eqs.
(26) and (30), together with the continuity of
$J\bigl(q(x)\bigr)$, that this combination must reduce to the
form
$$W(x,x')=J\bigl(q(x)\bigr){\delta}(x-x'),\eqno(33)$$
since the presence of derivatives of ${\delta}(x-x')$ would cause
the l.h.s. of Eq. (26) to diverge. On comparing Eqs. (18) and
(29), we see that
$$\bigl({\cal L}{\otimes}I+I{\otimes}{\cal L}'\bigr)
E\bigl({\xi}(x){\otimes}{\xi}(x')\bigr)=
2{\nabla}.\bigl(\bigl[K\bigl(q(x))J\bigl(q(x)\bigr)\bigr]_{sym}
{\nabla}{\delta}(x-x')\bigr).\eqno(34)$$
Hence, as $a{\otimes}b$ is the transpose of $b{\otimes}a$, 
$$\bigl[\bigl({\cal L}{\otimes}I\bigr)
E\bigl({\xi}(x){\otimes}{\xi}(x')\bigr)\bigr]_{sym}+
\bigl[\bigl({\cal L}'{\otimes}I\bigr)
E\bigl({\xi}(x'){\otimes}{\xi}(x)\bigr)\bigr]_{sym}=$$
$$2{\nabla}.\bigl(\bigl[K\bigl(q(x))J\bigl(q(x)\bigr)\bigr]_{sym}
{\nabla}{\delta}(x-x')\bigr),\eqno(35)$$
i.e., by Eqs. (30) and (33),
$$\bigl[{\cal L}\bigl(J\bigl(q(x)\bigr)
{\delta}(x-x')\bigr)\bigr]_{sym}
+\bigl[{\cal L}'\bigl(J\bigl(q(x')\bigr)
{\delta}(x'-x)\bigr)\bigr]_{sym}=$$
$$2{\nabla}.\bigl(\bigl[K\bigl(q(x))J\bigl(q(x)\bigr)\bigr]_{sym}
{\nabla}{\delta}(x-x')\bigr).\eqno(36)$$
We now infer from the formula for ${\cal L}$ given by Eq. (6),
together with Eq. (32), that 
$${\cal L}\bigl(J\bigl(q(x)\bigr){\delta}(x-x')\bigr)=$$
$${\Delta}\bigl[K\bigl(q(x)\bigr)J\bigl(q(x)\bigr)
{\delta}(x-x')\bigr]+
{\nabla}.\bigl[{\Psi}(q;x){\delta}(x-x')\bigr],\eqno(37)$$
while the interchange of $x$ and $x'$ in this formula yields the
equation 
$${\cal L}'\bigl(J\bigl(q(x')\bigr){\delta}(x'-x)\bigr)=
{\Delta}'\bigl[K\bigl(q(x')\bigr)J\bigl(q(x')\bigr)
{\delta}(x'-x)\bigr]+
{\nabla}'.\bigl[{\Psi}(q;x'){\delta}(x'-x)\bigr]=$$
$${\Delta}'\bigl[K\bigl(q(x)\bigr)J\bigl(q(x)\bigr)
{\delta}(x-x')\bigr]+
{\nabla}'.\bigl[{\Psi}(q;x){\delta}(x-x')\bigr]=$$
$$K\bigl(q(x)\bigr)J\bigl(q(x)\bigr){\Delta}{\delta}(x'-x)
-{\Psi}(q;x).{\nabla}{\delta}(x-x').\eqno(38)$$
It now follows easily from the last two equations, together with
the definition (31), that the
difference between the left and right hand sides of Eq. (36) is
simply ${\Phi}(q;x){\delta}(x-x')$. This signifies that the
assumption of short range correlations implies that
${\Phi}(q)=0$; or equivalently that a sufficient conditions for
the static ${\xi}$-correlations to be of long range is that
${\Phi}(q)$ does not vanish. 
\vskip 0.5cm
\centerline {\bf 5. Concluding Remarks}
\vskip 0.3cm
The quantum macrostatistical theory presented here is based on
the regression hypothesis together with the assumptions of local
equilibrium and chaotic current fluctuations. On this physical
basis, we have obtained both a generalised Onsager-Machlup
process and a generic picture of long range correlations in
nonequilibrium steady states for systems whose phenomenological
dynamics corresponds to a multi-component nonlinear diffusion.
Our derivation of these results depended on a coordination of the
spatial and temporal macroscopic scalings, which was facilitated
by the invariance of the assumed phenomenological law, given by
Eq. (1), under the scale transformation
$x{\rightarrow}{\lambda}x, \ t{\rightarrow}{\lambda}^{2}t$. We
remark that, for systems whose phenomenological laws do not have
any simple scale invariance, e.g. for Navier-Stokes
hydrodynamics, the situation is clearly much more complex; and
a generalisation of our results to such cases would presumably
require an intricate multi-scale analysis.  
\vskip 0.5cm
\centerline {\bf References}
\vskip 0.3cm\noindent
1. Bertini, L., De Sole, A., Gabrielli, D., Jona-Lasinio, G. and
Landim, C.: Macroscopic fluctuation theory for stationary
nonequilibrium states, J. Stat. Phys. {\bf 107} (2002), 635-675.
\vskip 0.2cm\noindent
2. Bonetto, R., Lebowitz, J. L. and Rey-Bellet, L.: Fourier's
law: a challenge to theorists: Pp. 128-150,
{\it Mathematical Physics 2000}, Ed. A. Fokas, A. Grigorian, T.
Kibble, B. Zegarlinski, Imperial College Press, 2000.
\vskip 0.2cm\noindent
3. Broidioi, M., Momont, B. and Verbeure, A.: Lie algebra of
anomolously scaled fluctuations: J. Math. Phys. {\bf 36} (1995),
6746-6757.
\vskip 0.2cm\noindent
4. Derrida, B., Lebowitz, J. L. amd Speer, E. R.: Large deviation
of the density profile in the symmetric simple exclusion process,
J. Stat. Phys. {\bf 107} (2002), 599-634.
\vskip 0.2cm\noindent
5. Dorfman, J. R., Kirkpatrick, T. R. and Sengers, J. V.: Generic
long range correlations in molecular fluids, Annu. Rev. Phys.
Chem. {\bf 45} (1994), 213-239.
\vskip 0.2cm\noindent
6. Gallavotti, G.: Chaotic hypothesis: Onsager reciprocity and
fluctuation-dissipation theorem: J. Stat. Phys. {\bf 84} (1996),
899-926.
\vskip 0.2cm\noindent
7. Goderis, D, Vets, P. and Verbeure, A.: Noncommutative central
limits, Prob. Th. Re. Fields {\bf 82} (1989), 527-544.
\vskip 0.2cm\noindent
8. Grinstein, G., Lee, D.-H. and Sachdev, S.: Conservation laws,
anisotropy and "self-organized criticality" in noisy
nonequilibrium systems, Phys. Rev. Lett. {\bf 64} (1990), 1927-
1930.
\vskip 0.2cm\noindent
9. Landau, L. D. and Lifschitz, E. M.: {\it Fluid Mechanics},
Pergamon, Oxford, 1984.
\vskip 0.2cm\noindent
10. Nelson, E.: {\it Dynamical Theories of Brownian Motion},
Princeton Univ. Press, Princeton, 1972. 
\vskip 0.2cm\noindent
11. Onsager, L.: Reciprocal relations in irreversible processes,
Phys. Rev. {\bf 37} (1931), 405-426; {\bf 38} (1931), 2265-2279.
\vskip 0.2cm\noindent
12. Onsager, L. and Machlup, S.: Fluctuations and irreversible
processes, Phys. Rev. {\bf 91} (1953), 1505-1512; {\it ibid}
1512-1515.
\vskip 0.2cm\noindent
13. Ruelle, D.: Positivity of entropy production in
nonequilibrium statistical mechanics, J. Stat. Phys. {\bf 85}
(1996), 1-25.
\vskip 0.2cm\noindent
14. Ruelle, D.: Natural nonequilibrium states in quantum
statistical mechanics, J. Stat. Phys. {\bf 98} (2000), 57-75.
\vskip 0.2cm\noindent
15. Schwartz, L.: {\it Th\'eorie des Distributions}, Tome 1,
Hermann, Paris, 1951. 
\vskip 0.2cm\noindent
16. Sewell, G. L.: Quantum macrostatistics and irreversible
thermodynmica, Pp. 368-83 of {\it Lecture Notes in
Mathematics} Vol. 1442, Ed. L.Accardi and W. Von Waldenfels,
Springer, Berlin, 1990.
\vskip 0.2cm\noindent
17. Sewell, G. L.: Towards a macrostatistical mechanics, Pp.
130-152 of {\it Mathematical Physics
towards the 21st Century}, Ed. R. N. Sen and A. Gersten, Ben-
Gurion University of the Negev Press, 1994.  
\vskip 0.2cm\noindent
18. Sewell, G. L.: {\it Quantum Mechanics and its Emergent
Macrophysics}, Princeton Univ. Press, Princeton, Oxford, 2002.
\vskip 0.2cm\noindent
19. Sewell, G. L.: In preparation.
\vskip 0.2cm\noindent
20. Spohn, H.: Long range correlations for stochastic lattice
gases in a nonequilibrium state, J. Phys. A {\bf 16} (1983),
4275-4291. 
\vskip 0.2cm\noindent
21. Streater, R. F. and Wightman, A. S.: {\it PCT, Spin and
Statistics, and All That}, W. A. Benjamin, New York, 1964.
\vskip 0.2cm\noindent
22. Tasaki, S. and Matsui, T.: Fluctuation theorem,
nonequilibrium steady states and the MacLennan-Zubarev ensembles
of $L^{1}$-asymptotical abelian $C^{\star}$-dynamical systems,
Pp. 100-119 of {\it Fundamental Aspects of Quantum Physics}, Ed.
L. Accardi and S. Tasaki, World Scientific, 2003; mp-arh 02-533
(2002).
\end